\documentclass[conference,a4paper]{IEEEtran}
\usepackage{multicol}
\usepackage[utf8]{inputenc}
\usepackage{footnote}
\usepackage[T1]{fontenc}
\usepackage{url}
\usepackage[ruled,vlined]{algorithm2e}
\usepackage{ifthen}
\usepackage{cite}
\usepackage{subfig}
\usepackage[cmex10]{amsmath}
\usepackage{comment}
\usepackage{amsthm}
\usepackage{amsmath,amssymb,amsfonts}
\usepackage{nicefrac}
\usepackage{algorithmic}
\usepackage{bm}
\usepackage{mathtools}
\usepackage{xcolor}

\usepackage{hyperref}
\usepackage[shortlabels]{enumitem}
\usepackage{cleveref}
\hypersetup{
    colorlinks=true,
    linkcolor=blue,
    filecolor=magenta,      
    urlcolor=cyan,
    pdftitle={Overleaf Example},
    pdfpagemode=FullScreen,
}

\definecolor{DarkGreen}{rgb}{0.1,0.5,0.1}
\definecolor{DarkRed}{rgb}{0.5,0.1,0.1}
\definecolor{DarkBlue}{rgb}{0.1,0.1,0.5}
\definecolor{DarkPurple}{rgb}{0.5,0.2,0.5}

\newtheorem{definition}{Definition}

\newtheorem{example}{Example}

\newcommand{\off}[1]{}

\title{Distributed Computations with Layered Resolution}

\author{%
   \IEEEauthorblockN{Homa Esfahanizadeh\IEEEauthorrefmark{1},
                     Alejandro Cohen\IEEEauthorrefmark{2},
                     Muriel M\'edard\IEEEauthorrefmark{1},
                     and Shlomo Shamai (Shitz)\IEEEauthorrefmark{2}}
    \IEEEauthorblockA{\IEEEauthorrefmark{1}%
                      RLE, MIT, Cambridge, MA, USA,  \{homaesf,medard\}@mit.edu}       
   \IEEEauthorblockA{\IEEEauthorrefmark{2}%
                      Faculty of Electrical and Computer Engineering, Technion, Israel, \{alecohen,sshlomo\}@technion.ac.il}

\vspace{-0.8cm}}

\begin{document}

\maketitle

\begin{abstract} 
Modern computationally-heavy applications are often time-sensitive, demanding distributed strategies to accelerate them. On the other hand, distributed computing suffers from the bottleneck of slow workers in practice. Distributed coded computing is an attractive solution that adds redundancy such that a subset of distributed computations suffices to obtain the final result. However, the final result is still either obtained within a desired time or not, and for the latter, the resources that are spent are wasted. In this paper, we introduce the novel concept of layered-resolution distributed coded computations such that lower resolutions of the final result are obtained from collective results of the workers -- at an earlier stage than the final result. This innovation makes it possible to have more effective deadline-based systems, since even if a computational job is terminated because of timing, an approximated version of the final result can be released. Based on our theoretical and empirical results, the average execution delay for the first resolution is notably smaller than the one for the final resolution. Moreover, the probability of meeting a deadline is one for the first resolution in a setting where the final resolution exceeds the deadline almost all the time, reducing the success rate of the systems with no layering.
\end{abstract} 

\section{Introduction}

A distributed computational system consists of multiple workers that can run as a single system to collaboratively respond to heavy computational jobs. Distributed computing offers major benefits such as scalability, parallelism, robustness, and cost efficiency. Delay has a major importance in distributed systems, as emerging computational applications demand a timely response to their computational jobs to make real-time automatic decision, e.g., autonomous driving and smart cities. In fact, often a computational response matters only if it is received within a reasonable time. On the other hand, the distributed computing settings are in practice heterogeneous with stochastic behaviour, making them susceptible for occasionally having large unexpected delays due to the stragglers. Therefore, relying on distributed resources to obtain timely results requires solutions that benefit from both coding and scheduling to add redundancy and to manage the heterogeneous environment. 

\begin{figure}
    \centering
     \includegraphics[trim=1.7cm 0cm 0cm 0cm,width=7.50cm,keepaspectratio]{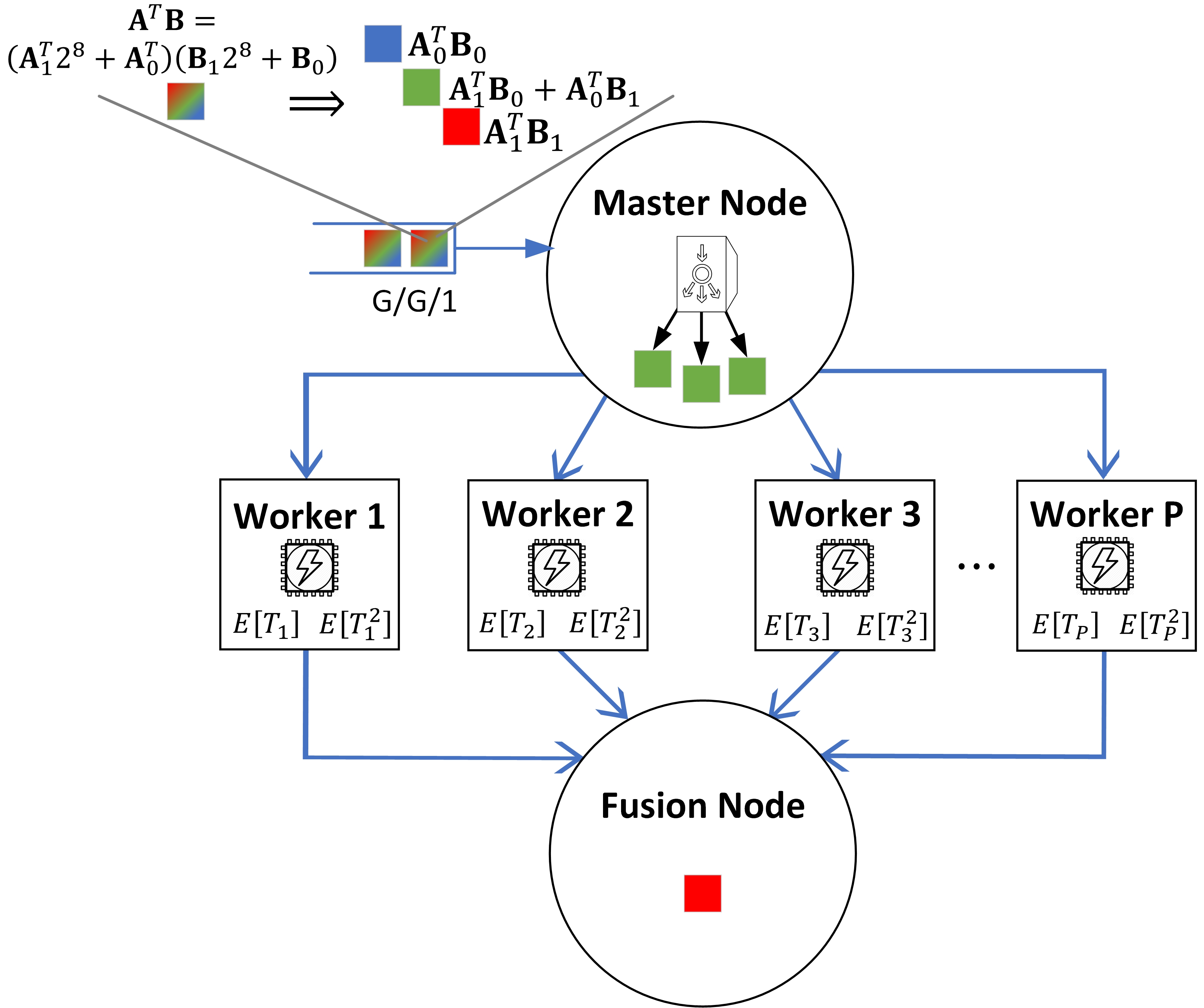}\vspace{-0.2cm}
    \caption{\small{Distributed coded matrix multiplication with layering: Operand matrices $\mathbf{A}$ and $\mathbf{B}$ have $16$-bit elements. The full precision outcome with $32$-bit elements, i.e., $\mathbf{A}^T \mathbf{B}=\mathbf{A}_1^T \mathbf{B}_12^{16}+(\mathbf{A}_1^T \mathbf{B}_0+\mathbf{A}_0^T \mathbf{B}_1)2^{8}+\mathbf{A}_0^T \mathbf{B}_0$, can be obtained in three layers of resolution.}\vspace{-0.7cm}}
   \label{fig:demo}
\end{figure}

Our problem setting consists of a master node equipped with a queue of arriving computational jobs, a heterogeneous cluster of worker nodes, and a fusion node. The master node serves the arriving jobs in order, splits their computational load into smaller tasks, and distributes these tasks among the workers. The workers individually work on their assigned tasks, and send the results of each task to the fusion node once ready. The fusion node aggregates the received task results and prepares the final result, Fig.~\ref{fig:demo}. 

The key requirement of having low-delay distributed systems has led to new advancements in (1) introducing redundant computations to mitigate the problem of slow workers (stragglers) and (2) utilizing informed scheduling to maximize the resource utilization. For the first trajectory, in distributed coded computation line of work, the computational load is encoded to include some redundancy, and the fusion node is able to obtain the final result of a computational job upon receiving a subset of the task results. Distributed coded computation has been studied for a variety of problems, e.g., matrix multiplication \cite{8002642,8437852,lee2017high,suh2017matrix,baharav2018straggler,yu2017polynomialn,Dutta2020,mallick2019fast,NIPS2017_e6c2dc3d}, gradient descent algorithm \cite{raviv2020gradient,dutta2019short,tandon2017gradient}, data shuffling, convolution, and fast Fourier transform \cite{song2019pliable,attia2019near,dutta2017coded,yang2016fault,jeong2018masterless,yu2017coded}, etc. For the second trajectory, nonuniform load balancing is considered to distribute the computational load among workers considering their various capabilities \cite{Reisizadeh,Avestimehr}. Recently, joint coding and scheduling was introduced to reduce end-to-end execution delay for a system with heterogeneity and stochastic response \cite{cohen2021stream,HomaInfocom}. The idea is to close the gap between distribution distance of variables that represent the response time of workers to their assigned coded tasks, via an optimized load balancing.

To the best of our knowledge, the previous distributed coded systems, while reducing the delay and uncertainty, consider only a single delay factor which indicates the time it takes from the job arrival until when the fusion node receives enough task results to release the final result. For data communications, layering has been already considered under multi-level delay or quality constraint using broadcast approach \cite{tajer2021broadcast,nikbakht2019mixed,nikbakht2020multiplexing,cohen2022broadcast,LimaMedard,KimMedard} and in the classical notion of distributed CEO problem \cite{berger1996ceo}, where delivery of different distortion (resolutions) is studied \cite{chen2008robust}. However, layering has been rarely investigated in the realm of distributed computations. 

In this paper, we propose a novel \textit{layered-resolution} version of distributed coded computations, where the fusion node can release an approximated result earlier than when releasing the final result is feasible, thus offering multiple delay factors -- one per resolution. To highlight the importance of this added feature, consider a system with a deadline, where if the computational time of a job exceeds a threshold and there are other jobs in the queue, the job will be terminated. In such a setting, without layering, all computational resources that are spent on the terminated job are wasted. However, using our scheme, an approximation of the final result can still be released for the terminated job.

According to our simulation results, the average delay of the first (lowest) resolution is notably lower than the average delay of the final (highest) resolution. Moreover, the success rate, i.e., the ratio of the number of successful job results to the total number of jobs, is almost one for the first resolution, in a setting where computing the final resolution is terminated quite often. This feature makes it possible to have more reliable and timely-manner computing solutions. 

\section{Problem Setting and Preliminaries}

Our distributed computing environment consists of a master node, a heterogeneous cluster of $P$ worker nodes, and a fusion node, see Fig.~\ref{fig:demo}. The master node serves the computational jobs according to their order of arrival. It encodes the computational load of each job into several smaller tasks (including some redundant tasks), and distributes the tasks among the workers. The parameter $\Omega\geq 1$, called redundancy ratio, indicates the ratio of the total number of tasks to the number of tasks needed for retrieving the final result. The workers send the result of each computational task to the fusion node as soon as ready. The fusion node can then obtain the final result upon receiving a sufficient subset of task results. The master node removes from the system the remaining tasks from the resolved job, called purging. It then encodes and distributes the computations related to the next job in its queue (if any).  

In this paper, we add a new dimension to our distributed coded computational setting, where the fusion node wishes to obtain lower resolutions of the final result, at an earlier time than when the final result is ready. This feature can be very helpful for the systems that have a deadline. This is because if the deadline is not met, still lower resolutions of the final response can be released. Our distributed system can be modeled with a G/G/1 queue, i.e., a system where job inter-arrival time has a general distribution and job service time (collaboratively performed by the workers) has another general distribution \cite{Bhat2015}. We also assume the first and second moments of the response time of each worker to its computational assignment are provided, 
and may be utilized to reduce the execution delay of different layers of resolution.

For simplicity, we assume the communication delays are negligible, and we target matrix-matrix multiplication as our example of computational jobs - supported by their importance as the building blocks of other algorithms such as machine learning training and inference. Let $\mathbf{A}$ and $\mathbf{B}$ be two large matrices with high-resolution elements that we are interested to find their multiplication $\mathbf{A}^T\mathbf{B}$ using a distributed, coded, and multi-resolution strategy. Without loss of generality, we assume their elements belong to $\text{GF}(q)$, and $q$ is prime. In the following subsections, we introduce some preliminary information  that will be used in the rest of the paper: (1) coded matrix multiplications and (2) nonuniform balancing of coded computations for heterogeneous computational environments.

\subsection{Coded Matrix Multiplication}

Several methods have been proposed recently in the literature for distributed coded computation of multiplying two large matrices, e.g., \cite{8002642,8437852,lee2017high,suh2017matrix,baharav2018straggler,yu2017polynomialn,Dutta2020,mallick2019fast,NIPS2017_e6c2dc3d}. In this subsection, we review the polynomial codes for distributing the job of matrix-matrix multiplication into several smaller tasks \cite{NIPS2017_e6c2dc3d}. However, the proposed layering method can be easily adapted to various coding schemes for matrix-matrix multiplications and beyond.

In polynomial coding scheme, a task is multiplication of two smaller matrices compared to the original matrices, and the final result can be obtained via a subset of task results, as follows: Each of two original matrices are first split into smaller sub-matrices, as
\[
\mathbf{A}{=}\left[\mathbf{A}^0|\mathbf{A}^{1}|\cdots|\mathbf{A}^{n_1-1}\right] \text{ and } \mathbf{B}{=}\left[\mathbf{B}^{0}|\mathbf{B}^{1}|\cdots|\mathbf{B}^{n_2-1}\right].
\]
Then, 
\[
\textbf{X}^{i}=f_i(\mathbf{A}^0,\dots,\mathbf{A}^{n_1-1}) \text{ and } \textbf{Y}^{i}=g_i(\mathbf{B}^{0},\dots,\mathbf{B}^{n_2-1}) 
\]
have the same size as partitioned sub-matrices of $\mathbf{A}$ and $\mathbf{B}$, respectively, and are computational inputs of the $i$-th task, i.e., $({\textbf{X}^{i}})^T\textbf{Y}^{i}$. The final result $\mathbf{A}^T\mathbf{B}$ can be obtained by collection of any $k=n_1n_2$ task results out of $k\Omega$ tasks results. This means for any $S\subset\{1,\dots,k\Omega\}$ such that $|S|=k$, there exists a decoding function $h$ such that $\textbf{A}^T\textbf{B}=h(\{(\textbf{X}^{i})^T\textbf{Y}^{i}\}_{i\in S})$. For more details regarding the encoding and decoding, refer to \cite{NIPS2017_e6c2dc3d}.

\subsection{Joint Scheduling-Coding Distributed Coded Computations}

In order to mitigate the straggling effects, we need to also manage the system heterogeneity via load balancing among workers based on their various capabilities. For maximizing resource utilization and reducing job execution delay as a result, the resources need to be utilized nonuniformly such that the response time of all workers to their assignments related to a job have similar distributions \cite{HomaInfocom}. Let $k$ and $\Omega$ be the number of necessary task results, and the redundancy ratio, respectively. The parameters $E[T_p]$ and $E[T_{p}^2]$ denote the first and second moments of the computation time of the $p$-th worker for one job. To minimize the execution delay, the number of assigned tasks for each worker $p\in\{1,\dots,P\}$ must be set as follows \cite{HomaInfocom}:
\begin{equation}\label{eq:nonuniform_split}
    \kappa_p=\frac{b_p}{2\gamma m_p^2}{\left(-1+\sqrt{1+\frac{4\gamma m_p^2\theta}{b_p^2}}\right)},
\end{equation}
where $m_p\triangleq E[T_p]$, $\sigma_p^2\triangleq{E[T_p^2]-E[T_p]^2}$, $b_p\triangleq m_p+\gamma\sigma_p^2$. The parameter $\gamma>0$ adjusts the relative importance of the first moment and the second moment on the distance measure of two distributions, and in our simulations, we set its value to $1$. The value of $\theta>0$ is set such that $\sum_{p=1}^{P}\kappa_p=k\Omega$. The obtained $\kappa_p$ using (\ref{eq:nonuniform_split}) has a real value. We then choose the closest integers to the optimal values such that $\sum_{p=1}^{P}\kappa_p{=}k\Omega$. We use this optimal load split to distribute the coded computations of each resolution layer among a set of workers. For more details, refer to \cite{HomaInfocom}.

\section{Layered Distributed Coded Computations}

Our proposed solution is based on partitioning the operand matrices twice: (1) partitioning each element of matrices $\mathbf{A}$ and $\mathbf{B}$ into several chunks, e.g., most significant bits (MSB) and least significant bits (LSB). By doing so, we obtain several matrix-matrix multiplications, called \textit{mini-jobs}, each serving to one layer of resolution; (2) partitioning and encoding the mini-jobs of each layer into several smaller matrices, as practiced in the conventional coded matrix multiplication. Our core idea is then to utilize the coded computation, in conjunction with an effective scheduling, to assemble the layered-resolution result from a subset of the task results. 

The computational jobs again arrive to the queue of a master node according to a general distribution. The master node serves the jobs according to their order of arrival as follows: It first converts a job into $L$ layers of resolution, such that the $l$-th layer consists of $J(l)$ mini-jobs, $l\in\{0,\dots,L-1\}$. Here, a mini-job is also a matrix-matrix multiplication with lower-resolution elements. It is then these mini jobs that are encoded and split among the workers, starting from the mini jobs of the first layer. After resolving all mini jobs of one resolution layer, the master node distributes the computations related to the next resolution of the same job or the first resolution of the next job (if any).


We start with a simple example that shows how multiplication of two scalars can be performed in three layers of resolution with no additional cost. Then, we extend this idea to finer granularity of resolution and for coded multiplication of two large high-resolution matrices. In the following simple example, we first reduce $\mathbf{A}$ and $\mathbf{B}$ to be scalars (i.e., size $1\times 1$), and we represent them with $\alpha$ and $\beta$, respectively.

\begin{example}
Let $\alpha$ and $\beta$ be two $16$-bit scalars, and the computational job be finding their multiplication $\alpha\beta$. We write the two variables as $\alpha=\alpha_1 2^8+\alpha_0$ and $\beta=\beta_1 2^8+\beta_0$, where $\alpha_1$, $\alpha_0$, $\beta_1$, and $\beta_0$ are at-most-$8$-bit scalars. Then
\begin{equation*}
    \alpha\beta=\alpha_1\beta_1 2^{16}+(\alpha_1\beta_0+\alpha_0\beta_1)2^{8}+\alpha_0\beta_0.
\end{equation*}
This multiplication can be done in three layers, each consists of multiplication(s) of two $8$-bit scalars: (a) $\alpha_1\beta_1$, (b) $\alpha_1\beta_0$ and $\alpha_0\beta_1$, and (c) $\alpha_0\beta_0$. We highlight two interesting observations: First, the order of these three layers matters. The first layer needs to be done first so that the results of the second layer increase the precision of computations. Second, the total computational complexity does not change by dividing the computations into these three layers. The complexity of multiplication of two scalars with $t$ bits is $\mathcal{O}(t^2)$. Therefore, both one-time pass and layered computations result in the computational complexity of $\mathcal{O}(256)$.
\end{example}

Now consider the following, more general, split of two scalars $\alpha$ and $\beta$ that are defined over a $q$-ary field into $m$ chunks, where $\alpha_i$ (resp., $\beta_i$) is the $i$-th chunk of the ordered constituent symbols of $\alpha$ (resp., $\beta$). Each chunk has size $d$ symbols and thus belongs to $\text{GF}(q^d)$:
\begin{equation*}
    \mathbf{M}_\alpha=\left[
         \alpha_{m-1},
         \dots, \alpha_0\right],\;\;
         \mathbf{M}_\beta=\left[\beta_{m-1},\dots,\beta_0
    \right]
\end{equation*}
Then, 
\[
\alpha=\sum_{i=0}^{m-1}\alpha_i.q^{id},\;\;\beta=\sum_{j=0}^{m-1}\beta_j.q^{jd}.
\]
A simple example is the case where $m=2$, $q=2$ (binary field), and $d=8$. Then, $\alpha_0$ (resp., $\beta_0$) is the MSBs and $\alpha_1$ (resp., $\beta_1$) is the LSBs of $\alpha$ (resp., $\beta$).
Now, consider the multiplication of the two scalars,
\begin{equation*}
    \alpha\beta=\sum_{i=0}^{m-1}\sum_{j=0}^{m-1}\alpha_i\beta_j q^{(i+j)d}.
\end{equation*}
If we skip those summations that are multiplied with lower powers of $q$, we get an approximation of the final result, with a lower complexity. 

Finally, we incorporate this layering into the coded matrix multiplication of two large matrices. Let partition constituent symbols of two matrices $\mathbf{A}$ and $\mathbf{B}$ over a $q$-ary field, i.e.,
\begin{equation*}
    \mathbf{A}=\sum_{i=0}^{m-1}\mathbf{A}_i q^{id},\;\;\mathbf{B}=\sum_{j=0}^{m-1}\mathbf{B}_j q^{jd}.
\end{equation*}
Similarly, $\mathbf{A}_i$s and $\mathbf{B}_j$s have elements that belong to $\text{GF}(q^d)$, and
\begin{equation*}
    \mathbf{A}^T\mathbf{B}=\sum_{i=0}^{m-1}\sum_{j=0}^{m-1}\mathbf{A}_i^T\mathbf{B}_j q^{(i+j)d}.
\end{equation*}
We define the $l$-th resolution of computing $\mathbf{A}^T\mathbf{B}$ in Definition~\ref{def:res}.
\begin{definition}[resolution]\label{def:res}
    The $l$-th resolution of computing $\mathbf{A}^T\mathbf{B}$, where $0\leq l\leq L-1$ and $L=2m-1$, is:
    \begin{equation*}
        {\mathbf{A}^T\mathbf{B}}|_{l\text{-th}}=\hspace{-1cm}\sum_{\{(i,j):(2m-2)-l\leq i+j\leq (2m-2),0\leq i,j\leq m-1 \}}\hspace{-1cm}(\mathbf{A}_i)^T\mathbf{B}_j q^{(i+j)d}.
    \end{equation*}
\end{definition}

Based on Definition~\ref{def:res}, the first resolution which corresponds to $l=0$, is computing $(\mathbf{A}_{m-1})^T\mathbf{B}_{m-1}q^{2m-2}$ and requires one matrix-matrix multiplication with elements in $\text{GF}(q^d)$. On the other end, the full resolution $l=L-1=2m-2$, is computing $\mathbf{A}^T\mathbf{B}$ which requires $m^2$ matrix-matrix multiplications with elements in $\text{GF}(q^d)$. Moreover, obtaining the $l$-th resolution from the $(l-1)$-th resolution requires
\begin{equation*}
\begin{split}
J(l)&=|\{(i,j):i+j=(2m-2)-l,0\leq i,j\leq m-1 \}|\\
&=\min\{l+1,2m-1-l\},
\end{split}
\end{equation*}
additional matrix-matrix multiplication(s), which stand for the mini-jobs of the $l$-th layer. Here, $J(0)=0$, and it can be verified that
\[
\sum_{l=0}^{L-1}J(l)=m^2,
\]
as expected.

\begin{figure*}
    \centering
    \vspace{-0.2cm}
    \subfloat[\centering]{{\includegraphics[width=7.0cm]{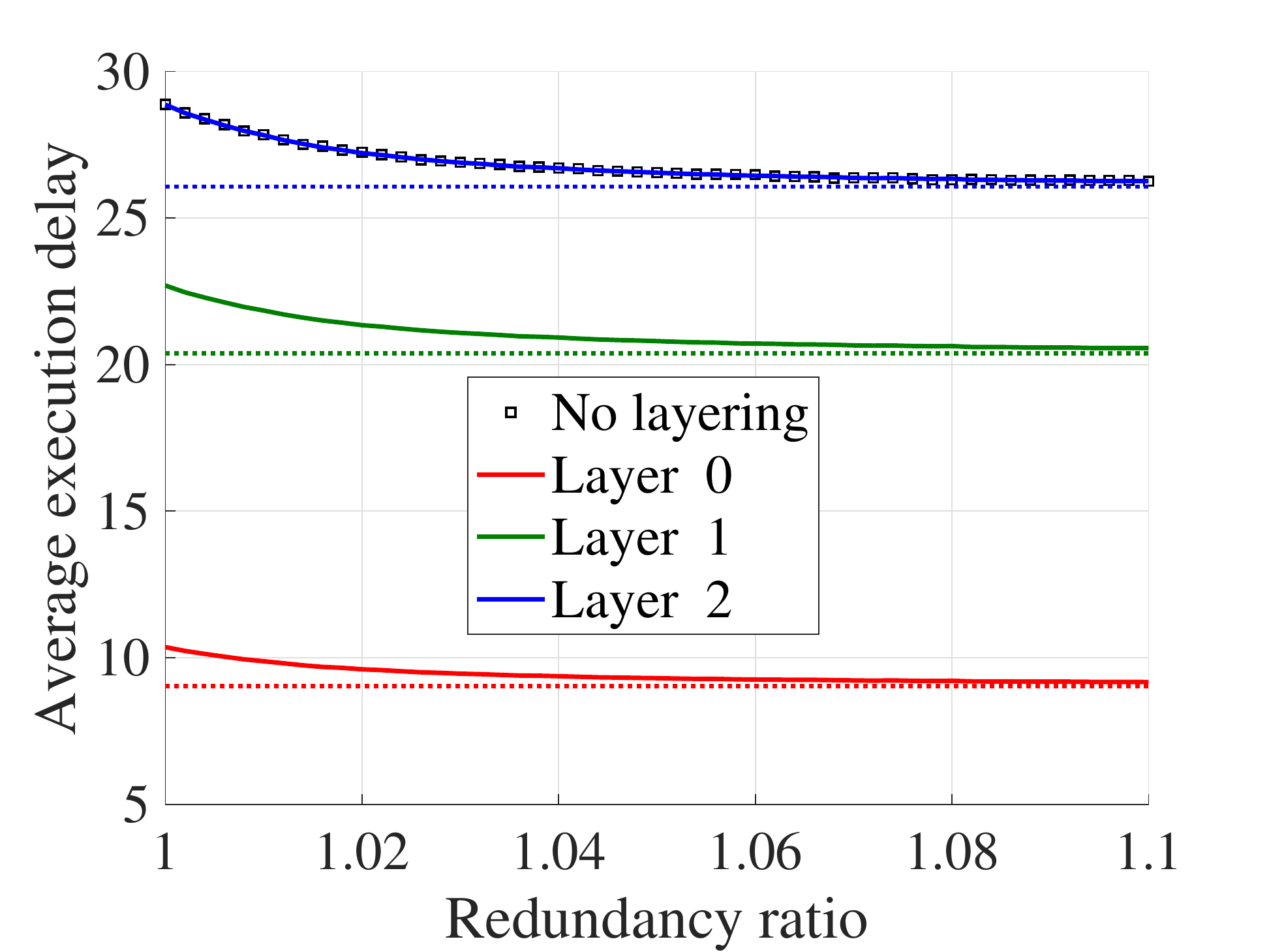} }}%
    \subfloat[\centering]{{\includegraphics[width=7.0cm]{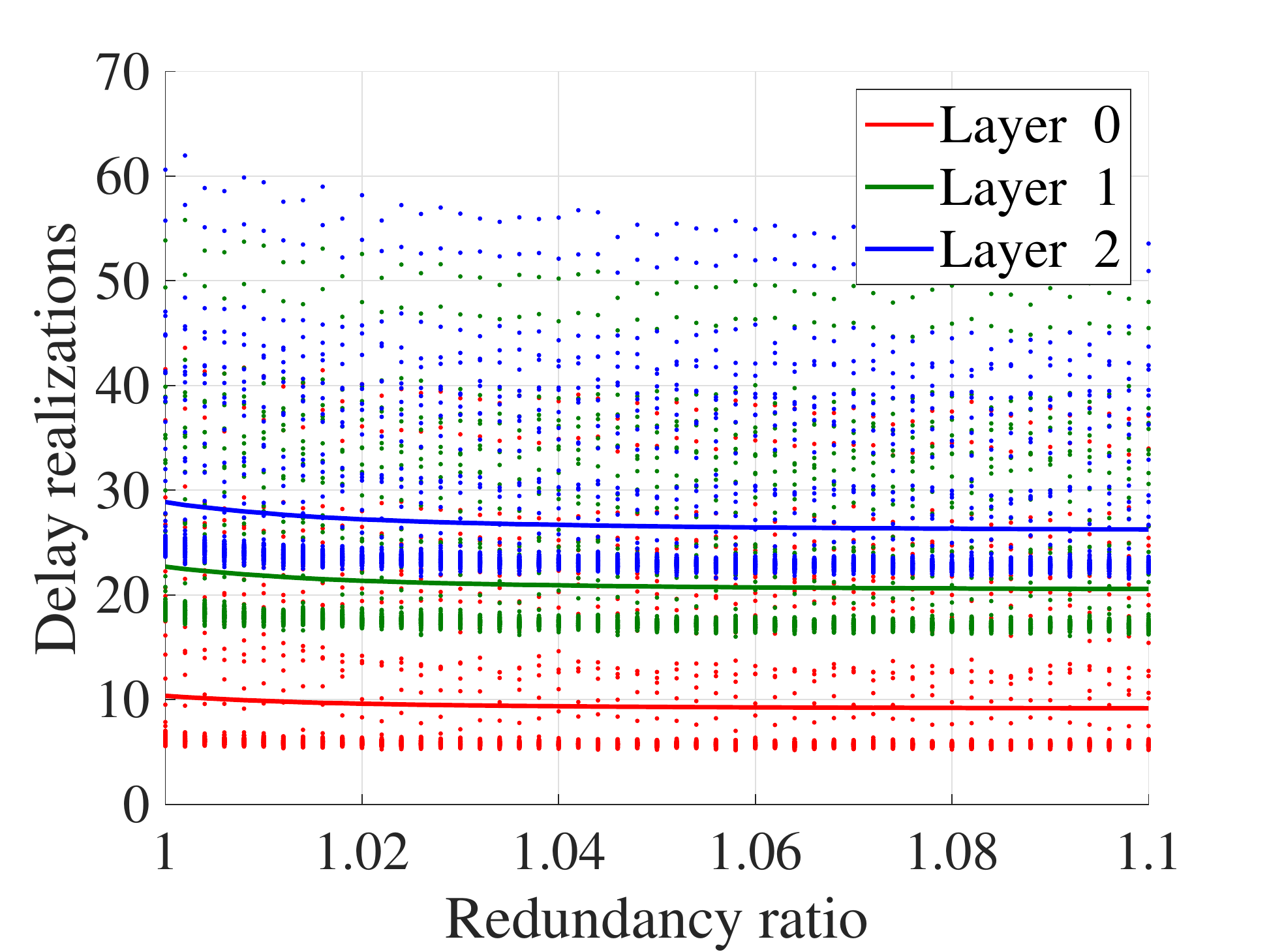} }}%
    \vspace{-0.2cm}
    \caption{\small{Delay analysis when the computational load is split into three layers vs. no layering: (a) average analysis along with a theoretically-driven lower bound, (b) realizations for $100$ jobs.}}
    \label{fig:delay_vs_omega}%
    \vspace{-0.4cm}
\end{figure*}

\subsection{Joint Coding and Scheduling with Layering}

The additional computations to upgrade the $(l-1)$-th resolution to the $l$-th resolution are equivalent to computing  $J(l)$ mini-jobs, i.e., matrix-matrix multiplications with low-resolution elements. The mini-jobs that contribute to each layer of resolution are then encoded into $k\Omega$ smaller matrix-matrix multiplication tasks, where $k$ is the number of  task results required by the fusion node to aggregate the final result for one matrix-matrix multiplication and $\Omega$ is the redundancy ratio. Next we define the delay profile of our layered computational solution in Definition~\ref{def:delayprofile}.
    
\begin{definition}[Delay profile]\label{def:delayprofile}
    We define execution delay of the $l$-th resolution of a job, denoted with $D(l)$, as the time the job arrives at the queue of the master node, until the time the fusion node is able to release the $l$-th resolution of the result.
\end{definition}
Since $D(l)$ varies from one job to another, due to the stochastic behaviour of system, we are interested in its distribution and expected value. Particularly, we are interested to obtain lower average delay for lower resolutions.

Our scheduling solution is for the master node to start with the first resolution, and encode and split the computational load of its mini-jobs (one-by-one) into several smaller tasks. Once one resolution is resolved (i.e., all its mini-jobs are finished), the master node will proceed with the next resolution of the same job or the first resolution of the next job in the queue (if any). We also occasionally consider a deadline, by which if the computational time of the job exceeds and there are other jobs in the queue, processing of the job will be terminated, and the last obtained resolution of its final result will be released by the fusion node.

\subsection{Theoretical Analysis on Delay Profile}

Let denote with $T_{p}$ the time it takes for the $p$-th worker to compute one complete job, and denote with $T_s$ the time it takes for all workers to collaboratively compute one job. The statistical parameters $E[T_p]$ and $E[T_p^2]$ are available, either by the designer or by tracking the workers' behaviour. Then, the service rate of the $p$-th worker is $1/E[T_p]$. A lower bound on $E[T_s]$ (the time it takes for the workers to finish the computations related to a job (from the time they start serving it) is obtained by approximating the whole system with just one worker whose service rate is summation of the service rates of all workers \cite{HomaInfocom}, i.e., 
\[
    E[T_s]\geq1/{\sum_{p=1}^{P}({1}/{E[T_p]}}). 
\]
For G/G/1 queuing model, with job inter-arrival time $T_a$, the average execution time (from job arrival to delivery and thus including the waiting time in the queue) is approximated by \cite{marchal1976approximate}:
\begin{equation}\label{eq:Kingmansformula}
    E[D]\approx E[T_s]+E[T_s]\left(\frac{\rho}{1-\rho}\frac{c_a^2+c_s^2}{2}
    \right).
\end{equation}
Here, 
\[
    \rho=E[T_s]/E[T_a], \quad c_a^2=(E[T_a^2]-E[T_a]^2)/E[T_a]^2,
\]
and 
\[
    c_s^2=(E[T_s^2]-E[T_s]^2)/E[T_s]^2.
\]
Incorporating the lower bound of $E[T_s]$ into (\ref{eq:Kingmansformula}) results in a lower bound for the average job delay for the distributed system with no layering. Here, the first part of the summation represents the average computational delay and the second part represents the average queuing delay.

When we also incorporate layering, the queuing delay is still the same, for a system with no job termination. However, the computational delay of lower layers are smaller. Let $T_s^l$ represent the time it takes for all workers to collaboratively compute the $l$-th resolution of the job. Since the load of computing $l$-th resolution requires $\sum_{i=0}^l J(i)$ mini-jobs, each having the same complexity, and there are $m^2$ total mini-jobs across all layers, we have \vspace{-0.2cm}
\begin{equation}\label{eq:LB2}
    E[T_s^l]\geq \frac{\sum_{i=0}^l J(i)}{m^2} \frac{1}{\sum_{p=1}^{P}\frac{1}{E[T_p]}}.\vspace{-0.2cm}
\end{equation}
Therefore, we have\vspace{-0.3cm}
\begin{equation}\label{eq:Kingmansformula2}
    E[D(l)]\approx E[T_s^l]+E[T_s]\left(\frac{\rho}{1-\rho}\frac{c_a^2+c_s^2}{2}
    \right).
\end{equation}

We demonstrate next, with our empirical results, that this is a tight lower bound, which is achieved by introducing slight computational redundancy per layer (for $\Omega\simeq 1.06$).

\section{Simulation Results}

We first describe the parameters of our system with $P=5$ workers: The job arrival has a Poisson distribution with rate $\lambda=0.01$. The number of critical tasks per matrix-matrix multiplication is $k=1000$. We assume the time it takes for the $p$-th worker to respond to an assignment with complexity $c$ has an exponential distribution with parameter $\mu_p/c$. Here, $\mu_p$ indicates the operation rate of the $p$-th worker, and the value of $\mu_p$ for the five workers used in this section are $[385.95,650.92, 373.40,415.75,373.98]$. We consider each element of operand matrices are partitioned into $m=2$ chunks, and thus we have $L=2m-1=3$ layers of resolution, see Fig~\ref{fig:demo}. The computational complexity of each task with no layering is set to $50$, and thus the computational complexity of each task when utilizing the layering mechanism is $12.5$.

\begin{figure*}
    \centering
    \vspace{-0.2cm}
    \subfloat[\centering]{\includegraphics[width=7.0cm,keepaspectratio]{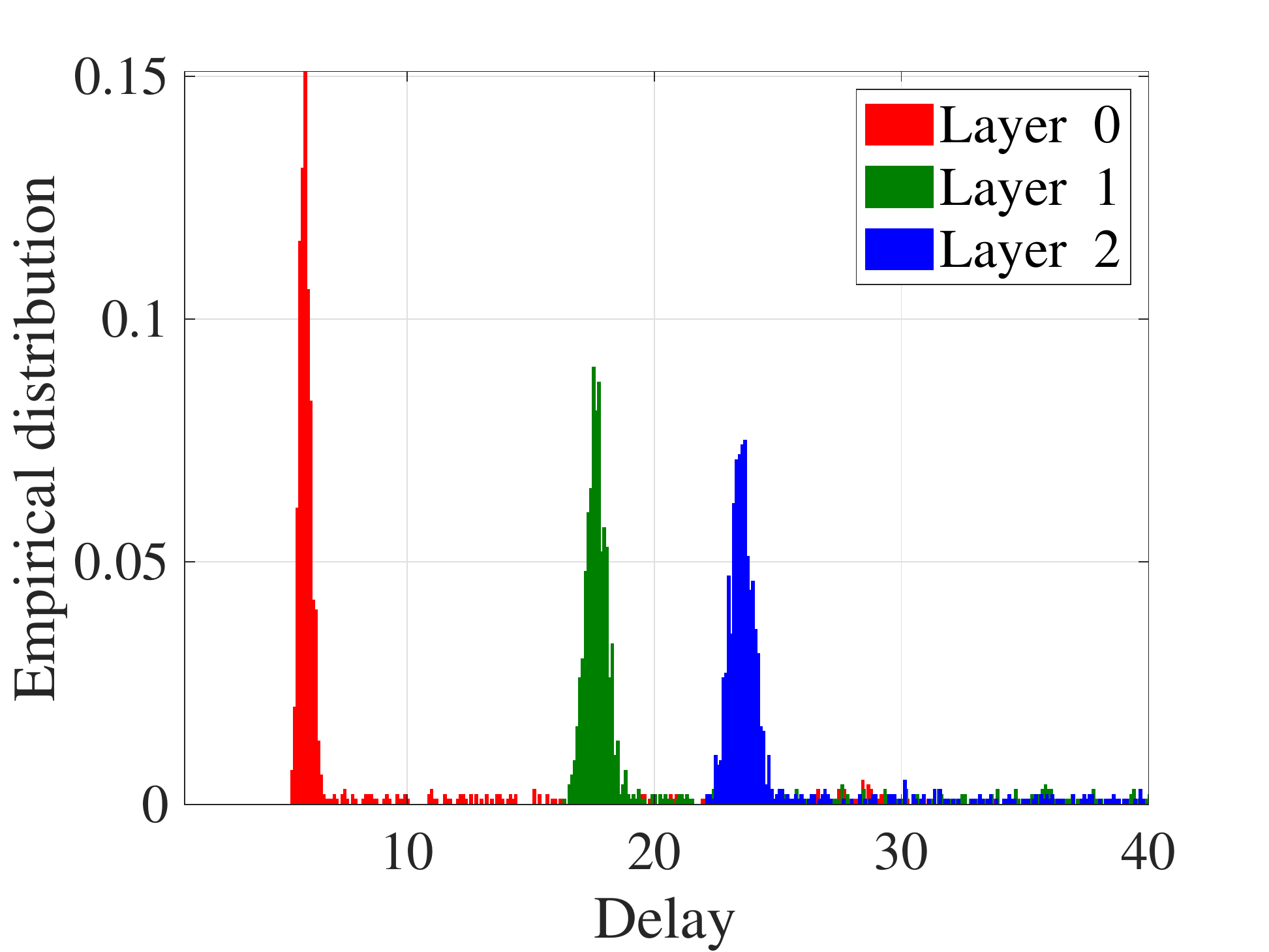}}
    \subfloat[\centering]{\includegraphics[width=7.0cm]{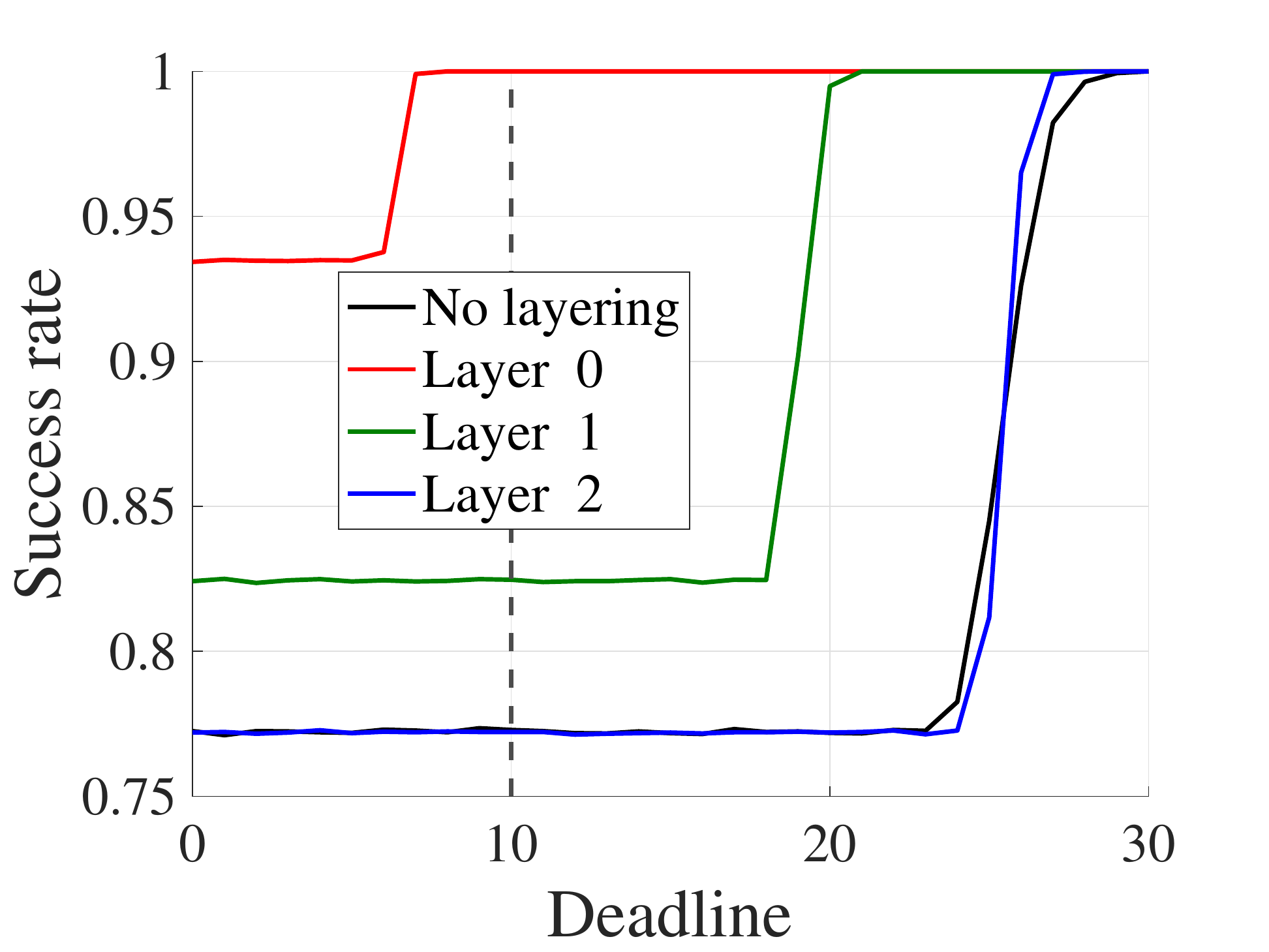}}
    \vspace{-0.2cm}
    \caption{\small{(a) Distribution and (b) success rate, for the execution delay of three layers of resolution, based on $1000$ job realizations and for the case of redundancy ratio $\Omega=1.018$.}}
    \label{fig:distribution}
    \vspace{-0.4cm}
\end{figure*}

Fig.~\ref{fig:delay_vs_omega}~(a) shows the average delay (arrival to delivery) versus redundancy ratio, obtained over the first $10000$ jobs in the queue, and Fig.~\ref{fig:delay_vs_omega}~(b) shows the delay realizations for the first $100$ jobs in the queue. Results are given for both settings, i.e., with and without layering. As we see, the last resolution has the average delay similar to the no-layering case. Thus, when we have the purging mechanism and when the communication delay is negligible, we can obtain an earlier lower-resolution versions of the final result almost at no additional cost. The difference between average delay of layer $l=0$ to $l=1$ is larger than $l=1$ to $l=2$. This is because upgrading the resolution to $l=1$ requires computing two mini-jobs while upgrading the resolution to $l=2$ requires one mini-job. Fig.~\ref{fig:delay_vs_omega}~(a) also shows the theoretically-driven lower bounds for average delay of each layer using our proposed formulation described in (\ref{eq:LB2}) and (\ref{eq:Kingmansformula2}). As seen, with about $6\%$ redundancy, those lower-bounds are empirically achievable across all resolution layers.

Fig~\ref{fig:distribution} (a) demonstrates the empirical distributions for the three layers of resolution. As seen, the higher layers have wider distributions because they accumulate the deviation from average behaviour of the earlier layers of resolution. However, the execution delays of the three layers still stand notably far from each other across major portion of realizations.

Finally, we perform an experiment where we impose a deadline to the system. The deadline specifies the maximum allowed computation time for each job when the system is busy. If the computational time of a job -- excluding the waiting time in the queue -- exceeds the deadline and there are subsequent job(s) in the queue, the job will be terminated. In the case of layering, lower resolutions of the job result might still be available although the job is terminated. The success rate is then defined as the ratio of the number of jobs, resp., certain resolution of jobs, that are finished (either because their computation time took less then the deadline or there was no other job in the queue) to the total number of jobs, for the first $1000$ jobs arrived at the queue.

Fig.~\ref{fig:distribution} (b) shows the success rate versus deadline value. As seen, the success rate is $1$ for the deadline value $10$, when the success rate for the higher resolutions or no-layering case are much lower. The success also depends on how busy the system is, as the termination criterion for a job is if its computation time exceeds both the deadline and the inter-arrival time to the next job in the queue. This experiment manifests that layering is a must-addition to distributed systems that have a deadline, to increase the effective resource utilization. Otherwise, the resources that are spent on terminated jobs will be completely wasted.

\off{\section{Conclusion}
We proposed a light-weight layering mechanism that can be incorporated into a variety of distributed coded computational schemes. The proposed layering allows lower resolutions of computational jobs be released at earlier stages than the full resolution. This added feature help to define a delay profile for distributed systems rather than a single delay indicator, and design more efficient and reliable systems.}


\bibliographystyle{IEEEtran}
\bibliography{IEEEabrv,references}
 
\end{document}